\documentclass[reprint,prb,aps,amsmath,amssymb,floatfix,superscriptaddress,longbibliography]{revtex4-1}
\usepackage{dcolumn}
\usepackage{bm}
\usepackage{graphicx}
\usepackage{amsthm}
\usepackage{color}
\usepackage{xcolor}
\usepackage{verbatim}
\usepackage{esint}
\usepackage[english]{babel}
\usepackage{amsfonts}
\usepackage{slashed}
\usepackage{latexsym}
\usepackage{bm}
\usepackage[colorlinks = true,
            linkcolor = red,
            urlcolor  = blue,
            citecolor = magenta,
            anchorcolor = red]{hyperref}
\usepackage{esint}
\usepackage{soul}
\usepackage{cancel}
\usepackage[normalem]{ulem}
\usepackage{empheq}
\usepackage{dsfont}
\usepackage{amsmath}
\usepackage{ulem}
\usepackage{epsfig}
\usepackage{mathrsfs}

\begin{document}
\title{Development of a Chip-Scale Optical Gyroscope with Weak Measurement Amplification Readout}
\author{Kagan Yanik}
\affiliation{Department of Physics and Astronomy, University of Rochester, Rochester, NY 14627, USA}
\affiliation{Institute for Quantum Studies, Chapman University, Orange, CA 92866, USA}
\author{Meiting Song}
\affiliation{Department of Electrical and Computer Engineering, University of California Santa Barbara, Santa Barbara, CA 93106, USA}
\affiliation{The Institute of Optics, University of Rochester, Rochester, NY 14627, USA}
\author{Yuhan Mei}
\affiliation{Department of Physics and Astronomy, University of Rochester, Rochester, NY 14627, USA}
\author{Jaime Cardenas}
\affiliation{The Institute of Optics, University of Rochester, Rochester, NY 14627, USA}
\affiliation{Department of Physics and Astronomy, University of Rochester, Rochester, NY 14627, USA}
\author{Andrew N. Jordan}
\affiliation{Institute for Quantum Studies, Chapman University, Orange, CA 92866, USA}
\affiliation{The Kennedy Chair in Physics, Chapman University, Orange, CA 92866, USA}
\affiliation{Schmid College of Science and Technology, Chapman University, Orange, CA 92866, USA}
\affiliation{Department of Physics and Astronomy, University of Rochester, Rochester, NY 14627, USA}

\date{\today}
	\begin{abstract}

The design of an integrated optical chip is proposed containing a rotation sensing ring resonator (optical gyroscope) coupled to an inverse weak value amplified Sagnac interferometer that amplifies the signal containing the phase information. We show that, for conservative parameter choices, our setup has a minimum detectable angular rotation rate $\sim0.1^{\circ}/$hr and an Allan deviation $\sim0.08^{\circ}/$hr under expected ideal conditions. We also show that for an appropriate amount of input power, our design can improve the signal-to-noise ratio, the precision of angular rotation rate, and error in detection by more than ten times compared to a Sagnac interferometer coupled to a ring resonator. 
	\end{abstract}
	\maketitle
	\section{Introduction}

Inertial rotation sensors have a variety of applications ranging from inertial navigation systems used in aircrafts to geophysical applications such as determination of astronomical latitude\cite{ 10.1007/978-3-540-39490-7_1}. Although mechanical gyroscopes were used as rotation sensors in the past, they have had disadvantages such as the presence of moving parts, warm up time and g-sensitivity\cite{ 10.1007/978-3-540-39490-7_1,venediktov2016passive}. The advent of the laser allowed ring resonators to be used as purely optical gyroscopes that do not contain the aforementioned disadvantages of mechanical gyroscopes\cite{chow1985ring,loukianov1999optical}. This is all thanks to the Sagnac effect that changes the optical length of the clockwise and counterclockwise paths around the ring such that they are not equal anymore\cite{post1967sagnac,10.1007/978-3-540-39490-7_1, passiveringres, Meyer:83, sagnac}. 

Weak value amplification (WVA) can amplify small parameters while suppressing certain technical noises such that the system yields an increased signal-to-noise ratio (SNR) and shot noise limited sensitivity\cite{feizpour2011amplifying,song2021enhanced,steinmetz2022enhanced,lyons2018noise,steinmetz2019precision}. WVA consists of three steps: (i) pre-selection of the initial state of the system, (ii) a weak perturbation to the system state, and (iii) post-selection of the final state of the system, which is nearly orthogonal to the initial state\cite{aharonov1988result,steinmetz2022enhanced,steinmetz2019precision}. Technical advantages of WVA were discussed in detail in Refs.\cite{jordan2014technical,dixon2009ultrasensitive}. 

In this paper, we discuss an integrated optical chip on which an optical ring resonator is coupled to an inverse weak value amplified (IWVA) Sagnac interferometer, which contains phase front tilters, a multi-mode directional coupler, and a multi-mode interferometer at the readout, as discussed in Refs.\cite{steinmetz2022enhanced,steinmetz2019precision,song2021enhanced}. While the optical ring resonator functions as an optical gyroscope that provides a phase shift due to rotation of the ring explained by the Sagnac effect, the IWVA Sagnac interferometer inverse weak value amplifies (IWVA), discussed below, the phase shift through phase front tilting. 
A minimum detectable angular velocity of the order of $10^{\circ}/$hr is a key challenge for miniaturized gyroscopes and motivates the research efforts in the field\cite{dell2014recent}. We show that, because of IWVA, our setup has a minimum detectable angular rotation rate $\sim0.1^{\circ}/$hr under expected ideal conditions and improves the SNR and phase resolution compared to a common Sagnac interferometer. It is worth distinguishing WVA and IWVA in the context of this work. WVA is used in measuring the spatial phase front tilt by using known phase shift\cite{song2021enhanced,hosten2008observation}. On the other hand, IWVA, which is the method used in this paper, is used in measuring the phase shift with amplified signal by using the known spatial phase front tilt\cite{song2021enhanced,starling2010precision}.

The paper is organized as follows: in Sec.~\ref{sec2}, we discuss the setup of our system, including the 50/50 beam splitter, the coupling between IWVA Sagnac interferometer and the optical ring, and the phase and amplitude amplification of the electric field coming out of the optical ring. In Sec.~\ref{results}, we discuss the 50/50 beam splitter before the output and derive the signal-to-noise ratio (SNR), phase resolution, and Allan deviation for two different systems: (i) a common Sagnac interferometer coupled to an optical ring and (ii) IWVA Sagnac interferometer coupled to an optical ring. We then compare the two systems and discuss the advantages of using IWVA Sagnac over common Sagnac interferometer. Finally, in Sec.~\ref{conc}, we conclude our results.

\section{The System}\label{sec2}
\begin{figure}[hbt!]
\centering
\includegraphics[scale=0.47]{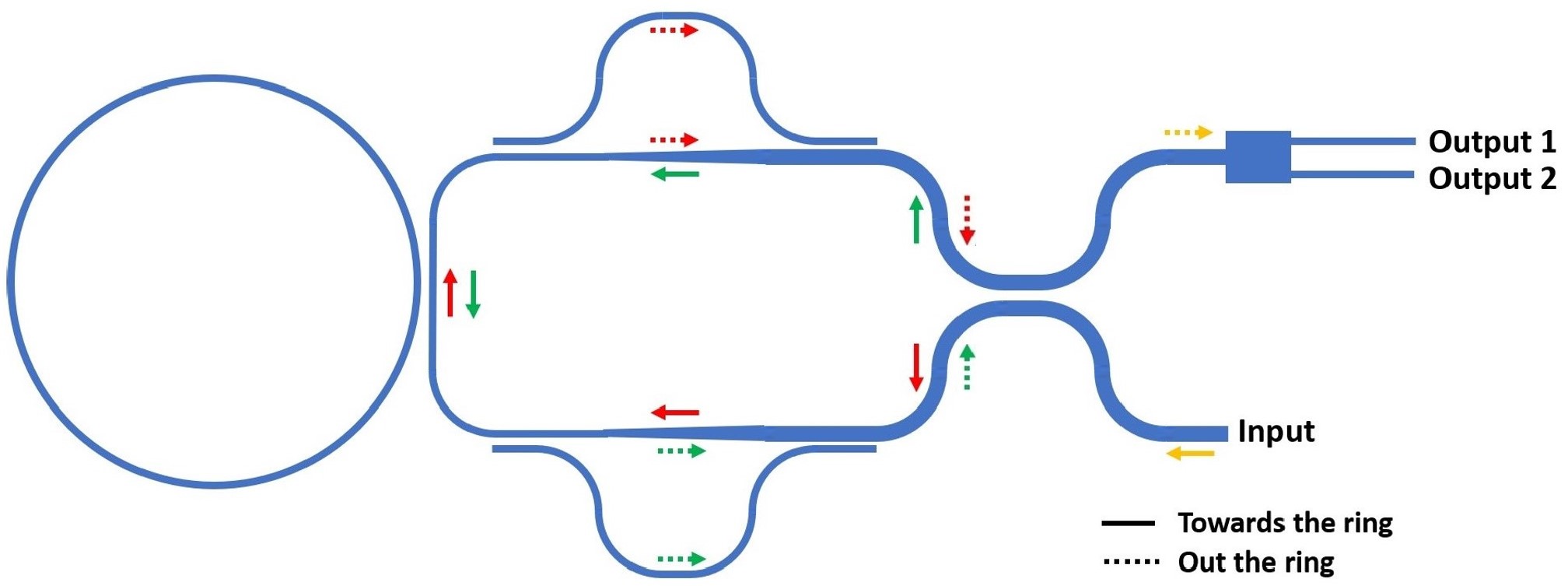} (a)\\
\includegraphics[scale=0.55]{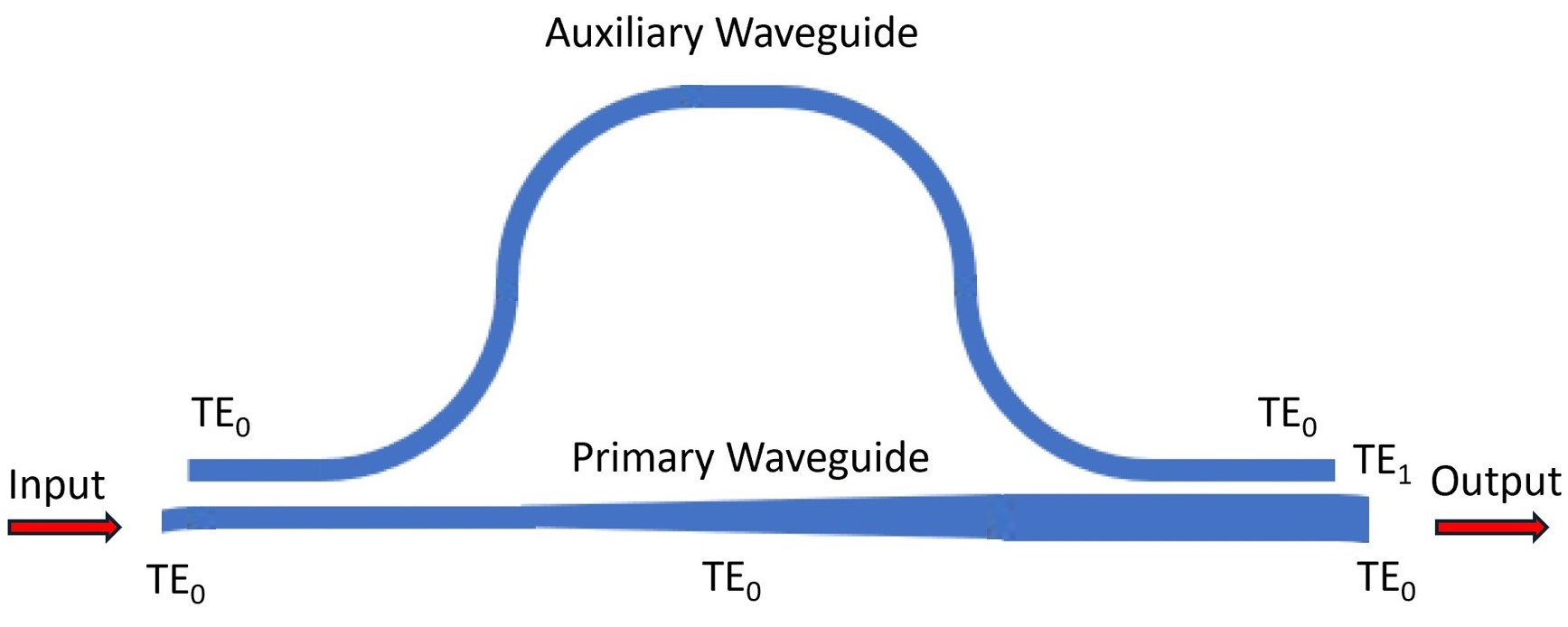} (b)\\
\includegraphics[scale=0.6]{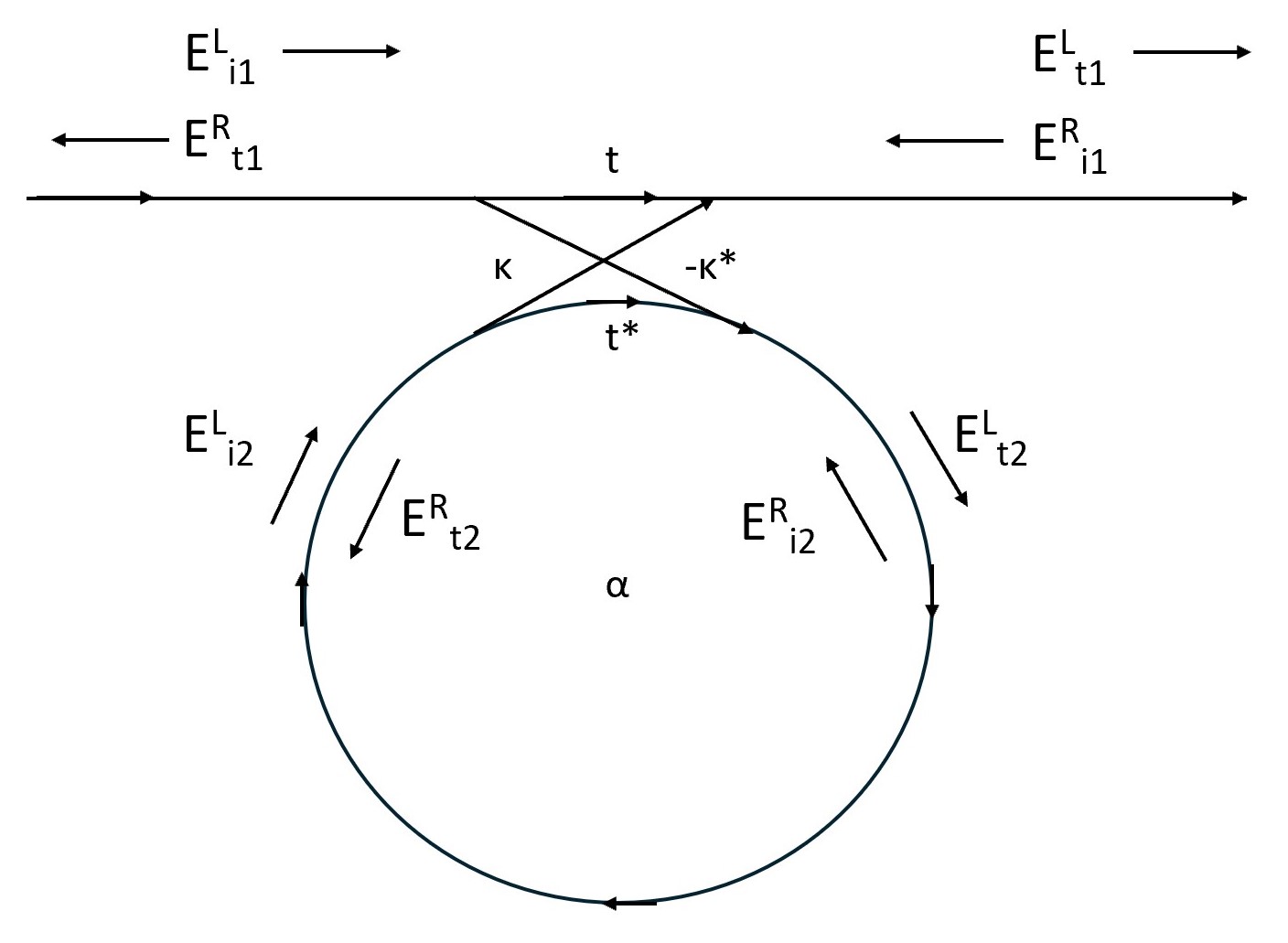}\\
(c)\\
 \caption{$\bold{a.}$ The considered device geometry (not to scale). The rotation sensing ring (left) is coupled with a single point of contact to the interferometric readout, based on the inverse weak value physics. Solid (dashed) arrows represent the input (output) field\cite{meitingthesis}. $\bold{b.}$ Spatial phase front tilter that couples a portion of $\rm TE_0$ mode to $\rm TE_1$ mode to create spatial phase front tilt in waveguides. $\bold{c.}$ Diagram of the scattering description of the ring. $\alpha$ accounts for the loss per cycle around the ring. $t$ ($t^{*}$) represents the portion of the electric field in IWVA Sagnac interferometer (optical ring) that is transmitted without interacting with the optical ring (IWVA Sagnac interferometer). $\kappa$ ($-\kappa^{*}$) represents the portion of the electric field that is transferred from the optical ring (IWVA Sagnac interferometer) to IWVA Sagnac interferometer (optical ring)\cite{yariv2000universal}.}
\label{fig:setup}
\end{figure}

Our system, shown in Fig.~\ref{fig:setup}{\color{red}a}, consists of a rotation sensing ring coupled with an IWVA Sagnac interferometer supporting $\rm TE_0$ and $\rm TE_1$ modes. The interferometer implements a version of inverse weak value amplification using phase front tilters and a multimode 50/50 beam splitter. The setup of this interferometer is discussed in detail in Refs.\cite{steinmetz2022enhanced,steinmetz2019precision,song2021enhanced}. The overall process of the inverse weak value amplification (IWVA) is as follows: the light is injected into the system through the input port on the bottom right of Fig.~\ref{fig:setup}{\color{red}a}. Then it goes through a multimode 50/50 beam splitter implemented by waveguide coupling (the point where the two waveguides, on the left of the input and output ports, are very close to each other). Hence, the two beams obtained after the beam splitter go through the left and right arms (lower and upper waveguides respectively in Fig.~\ref{fig:setup}{\color{red}a}). As they propagate through these arms, they narrow in transverse profile as the waveguides narrow down in width adiabatically to single ${\rm TE}_0$ mode waveguide. Since the IWVA Sagnac interferometer is coupled with the optical ring, the two beams enter the ring from opposite directions and circulate through the ring independently. Throughout these circulations, the optical ring will carry the wavelengths that match its resonance condition. On the other hand, the radiation coming out of the ring carries change in amplitude and phase and, most importantly, carries information about the rotation of the ring resonator. This is caused by the fact that the optical length of the two opposite paths around the ring are no longer equal due to rotation, as described by the Sagnac effect\cite{passiveringres}. In this work, we assume that the optical ring has a linear refractive index and, therefore, does not have any nonlinear behavior. The beams coming out of the optical ring then propagate through the spatial phase front tilters discussed in Ref.~\cite{song2021enhanced} and shown in Fig.~\ref{fig:setup}{\color{red}b}. The spatial phase front tilters operate as follows: the primary and auxiliary waveguides are initially identical single mode waveguides carrying $\rm TE_0$ mode. The primary waveguide then couples a small portion of the beam in the $\rm TE_0$ mode of the auxiliary waveguide. Then the width of the primary waveguide is increased adiabatically to make it a multimode waveguide that can support both $\rm TE_0$ and $\rm TE_1$ modes. The light in the $\rm TE_0$ mode in the primary waveguide stays in $\rm TE_0$ mode since the waveguide is widened adiabatically. Moreover, the widening of the primary waveguide is designed precisely so that its $\rm TE_1$ mode supported after the taper is phase matched to the $\rm TE_0$ mode in the auxiliary waveguide. Consequently, the light in the auxiliary waveguide couples back into the tapered primary waveguide in $\rm TE_1$ mode, giving us a combination of $\rm TE_0$ and $\rm TE_1$ modes at the end of the primary waveguide\cite{song2021enhanced,luo2014wdm,crespi2011integrated}, as shown in Fig.~\ref{fig:setup}{\color{red}b}. After the light beams, containing both $\rm TE_0$ and $\rm TE_1$ modes, go through the 50/50 beam splitter, we obtain two outputs: one with high intensity light that carries little information about the ring's rotation (bright port) and the other with lower intensity that carries most of the information about the rotation of the ring (dark port). Finally, at the dark port, light goes through a multimode interfering (MMI) region, the blue box at the top right of Fig.~\ref{fig:setup}{\color{red}a}, with two outputs. The optical power difference between the two outputs of the MMI depend on the ratio between the input $\rm TE_0$ and $\rm TE_1$ modes and contains the desired phase signal\cite{song2021enhanced}. We will be discussing these processes in more detail in the following subsections.

\subsection{Input}\label{sub:input}
To characterize the light entering and propagating through the system, we solve the frequency space Helmholtz equation that gives the mode structure of the traveling electromagnetic fields\cite{1130282273250302336,griffiths2021introduction}
\begin{equation}
    \nabla^2\bold{E}+n(\omega)^2k_0^2\bold{E}=0,
    \label{eq:helm}
\end{equation}
where $\bold{E}$ is the time and space dependent electric field, $k_0$ is the wavenumber of the light,
and $n(\omega)$ is the frequency-dependent index of refraction. For this setup, we assume the interferometer to be made of rectangular waveguides where the core exists in space for $|x|\leq d$ and $|y|\leq b$, and has a refractive index larger than the cladding layer. The electric field propagates in z direction. We assume $b\gg d$ so that the transverse electric field amplitude is practically independent of the y direction and is expressed as
\begin{equation}
    \bold{E}=E_0(x)e^{i\beta_k z}\hat{y},
\end{equation}
where $\beta_k$ is the frequency-dependent wavevector for mode $k$ of the electric field with frequency $\omega_k$. Assuming the time dependence is suppressed, only the zeroth and first order transverse electric modes (${\rm TE_0}(x)$ and ${\rm TE_1}(x)$) are required for weak value amplification with integrated
photonic devices\cite{song2021enhanced}.  

\subsection{50/50 Beam Splitter}
The input electric field at $\rm TE_0$ mode is injected through the left waveguide (the bottom waveguide in the right hand side of the Fig.~\ref{fig:setup}{\color{red}a}). Then it goes through a 50/50 beam splitter which is achieved by a directional coupler where the left and right (the top waveguide in the right hand side of the Fig.~\ref{fig:setup}{\color{red}a}) waveguides are brought very close to each other so that their electric field modes are coupled and power can transfer periodically between them\cite{introfiber}. Then the total electric field for the coupled left and right waveguides can be expressed as
\begin{equation}
    E_{total}=A(z)E^L(x)e^{i\beta_L z}+B(z)E^R(x)e^{i\beta_R z},
\end{equation}
where $E_{L,R}$ are the electric field modes of the waveguides, $\beta_{L,R}$ are the wave propagation speeds of the left and right waveguides, and $A$ and $B$ are the normalized amplitudes of the modes in left and right waveguides respectively\cite{huang1994coupled}. We assume the modes in the individual waveguides are unperturbed. The generalized coupling constant between any two waveguides $i$ and $j$ can be expressed as\cite{hiremath2005coupled,introfiber,haus1987coupled,huang1994coupled} 
\begin{equation}
    \kappa^{\rm DC}_{ij}=\frac{\omega_0\epsilon_0\int_{-\infty}^{\infty}\int_{-\infty}^{\infty} \Delta n_j^2 \bold{E}^{i*}\cdot\bold{E}^j \,dxdy }{\int_{-\infty}^{\infty}\int_{-\infty}^{\infty} \hat{z}\cdot(\bold{E}^{i*}\times\bold{H}^i+\bold{E}^i\times\bold{H}^{i*}) \,dxdy},
    \label{eq:couplingconstant}
\end{equation}
where $\bold{E}^i$ and $\bold{H}^i$ are the transverse electric and magnetic fields in waveguide $i$ respectively, $\omega_0$ is the angular frequency of the field mode, $\epsilon_0$ is the vacuum permittivity, $\Delta n_j^2=n^2(x)-n_j^2(x)$ with $n(x)$ being the refractive index of the two coupled waveguides and $n_j^2(x)$ being the refractive index of the isolated waveguide $j$. Based on the boundary conditions, $A(z)=\rm cos(\kappa^{\rm DC}_0z)$ and $B(z)=i\rm sin(\kappa^{\rm DC}_0z)$, where $\kappa^{\rm DC}_0=\kappa^{\rm DC}_{LR}=\kappa^{\rm DC}_{RL}$\cite{hiremath2005coupled,introfiber,haus1987coupled,huang1994coupled}. For a 50/50 beam splitter, we then need the length of the directional coupler to be $S=m\pi/(4\kappa^{\rm DC}_0)$, where $m$ is any odd integer. We can take $m=1$ for now. Then, the electric field coming outside the 50/50 beam splitter is $A(S)E^L(x,z)={\rm TE_0}(x)e^{i\beta_0z}/\sqrt{2}$ in the left arm, and $B(S)E^R(x,z)=i{\rm TE_0}(x)e^{i\beta_0z}/\sqrt{2}$ in the right arm for $z\geq S$. 

We can also express the outgoing electric fields with the transformation matrix that represents the beam splitter interactions\cite{yariv2000universal}
\begin{gather}
 \begin{bmatrix} E_{t1} \\ E_{t2} \end{bmatrix}
 =
  \begin{bmatrix}
   t &
   \kappa \\
   -\kappa^{*} &
   t^{*} 
   \end{bmatrix}
   \begin{bmatrix} E_{i1} \\ E_{i2} \end{bmatrix},
   \label{matrix bs}
\end{gather}
where $t$ is transmittance, $\kappa$ is reflectance and in a lossless system $|t|^2+|\kappa|^2=1$. Moreover, $E_{i1}$ and $E_{i2}$ are the input electric fields incident on the beam splitter from different input ports whereas $E_{t1}$ and $E_{t2}$ are the output electric fields from the beam splitter that are linear combinations of the two input fields.

\subsection{Coupling to the Optical Ring}\label{subsec:C}
After the two beams propagate through the beam splitter, they travel along the left and right waveguides and narrow in transverse profile as the waveguides gradually decrease in width. Then these beams are incident on the single-mode optical ring from opposite directions through point of contact coupling. Inside the ring, the beams independently circulate through the ring and interfere with themselves. The wavelength in resonance with $\rm TE_0$ mode is transferred to the ring, accumulates a Sagnac phase, and exits the ring and propagates through the waveguides to be inverse weak value amplified, which we will discuss in next subsection. The geometry of the coupling with the ring is shown in Fig.~\ref{fig:setup}{\color{red}c}. We can approach the coupling of the IWVA Sagnac interferometer and the ring resonator like a beam splitter expressed with the matrix in Eq.~(\ref{matrix bs}), where $t$ is the transmission amplitude through the coupling region and $\kappa$ is the coupling coefficient between the two waveguides\cite{yariv2000universal}. Assuming a lossless coupling, we can say that $|t|^2+|\kappa|^2=1$. Here, $E_{i1}$ and $E_{t1}$ are the input and output electric fields of the bus waveguide respectively. Similarly, $E_{i2}$ and $E_{t2}$ are the input and output electric fields of the ring resonator respectively. In addition to the beam splitter type relation that describes the geometry of this coupling, we also consider the boundary condition that reveals the relation between the input and output fields of the ring resonator:
\begin{equation}
    E_{i2}=\alpha e^{i(\theta\pm\phi)}E_{t2},
    \label{eq:BC}
\end{equation}
where $\alpha$ accounts for the loss per cycle around the ring and is expressed as $\alpha=e^{-\Gamma L/2}$ where $\Gamma$ is the power loss per unit length. $\theta=2\pi L/\lambda$ gives the geometric phase shift ($L=2\pi R$ is the circumference of the ring, $R$ is the radius of the ring and $\lambda$ is the wavelength of the light), and $\phi=2\pi\ell/\lambda$ gives the phase shift of light traveling in one direction for a single roundtrip due to Sagnac effect where $\ell=2A\Omega/c$ ($A$ is the area of the ring, $\Omega$ is the angular rotation rate, and $c$ is the speed of light). The Sagnac phase shift has opposite signs for counterpropagating directions. Under the assumption that there are no nonlinear effects present inside the ring, the phase shift $\phi$ due to Sagnac effect is assumed to be linear only. The plus or minus sign on the phase $\phi$ in Eq.~(\ref{eq:BC}) depends on whether the direction of the light propagation is with or against the rotation of the ring. Applying this boundary condition to the beam splitter type relation, we can solve for the relation between the input and output electric fields of the linear waveguide for both possible directions of the input field (left or right)

\begin{equation}
    E_{t1}^{L,R}=\frac{-\alpha+te^{-i(\theta\pm\phi)}}{-t^{*}\alpha+e^{-i(\theta\pm\phi)}} E_{i1}^{L,R}.
    \label{eq:electricfield}
\end{equation}
We assume that $\alpha$ and $t$ are real, and there is no rotation ($\phi=0$). We then expand ${\rm cos(\theta)}\approx1-(\theta-\theta_n)^2/2$ near the resonance condition ($\theta=\theta_n$), where $\theta_n=2\pi L/\lambda_n$. Here, $L$ is the circumference of the ring for which the resonant wavelengths ($\lambda_n$) satisfy $L=n\lambda_n$, where $n$ is the mode number and an integer. The power coming out of the ring is expressed as
\begin{equation}
    P_{out}=\frac{(\alpha-t)^2+\alpha t(\theta-\theta_n)^2}{(1-\alpha t)^2+\alpha t(\theta-\theta_n)^2}P_{in},
    \label{eq:pout_original}
\end{equation}
which has an inverse Lorentzian shape\cite{christopoulos2019calculation}. At resonance ($\theta=\theta_0$), it is then expressed as
\begin{equation}
    P_{out}=\bigg(\frac{\alpha-t}{1-\alpha t}\bigg)^2P_{in},
    \label{eq:Pout}
\end{equation}
where the input power is lost by a factor of $(\alpha- t)^2/(1-\alpha t)^2$ after coming out of the ring resonator. We notice that at the critical coupling limit ($\alpha=t$) and resonance condition ($\theta=\theta_n$), $P_{out}=0$. Therefore, all the incident light is transferred to the ring and no power is transmitted back to the IWVA Sagnac interferometer.    

\subsubsection{The Quality Factor}
\begin{figure}[!htb]
\centering
 \includegraphics[scale=0.65]{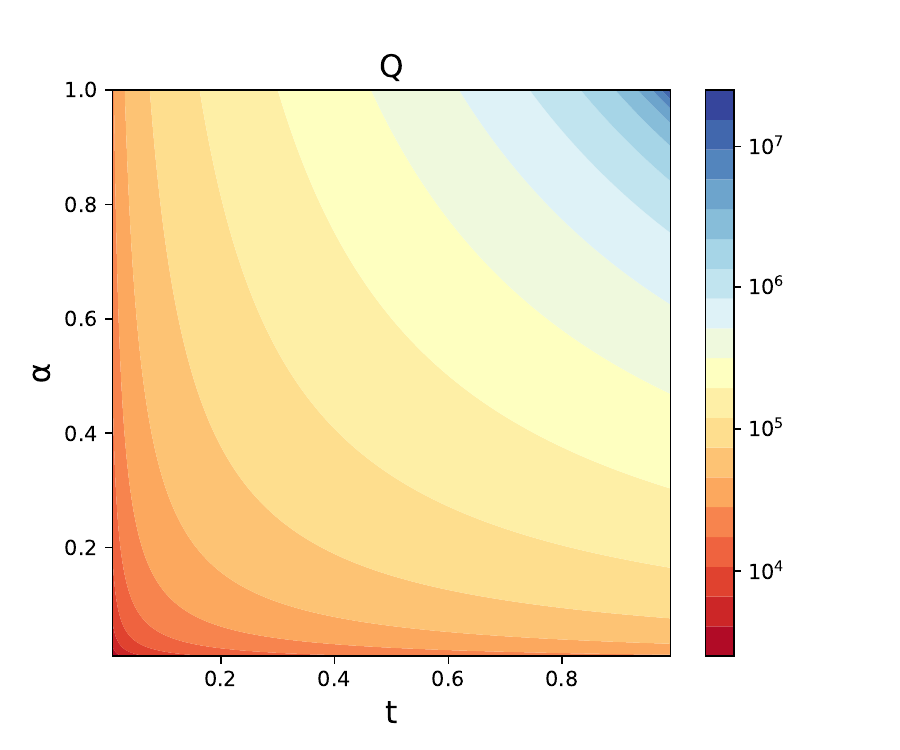}
 \caption{Contour plot of the loaded quality factor Q as a function of $\alpha$ and $t$. The radius of the ring is $R=10$mm and the resonant wavelength is $\lambda_n=1550$nm.}
 \label{fig:Qcontour}
\end{figure}

In light of these relations, we can determine the quality factor ($Q=\omega_n/\Delta\omega$) which is defined as the ratio between the resonance frequency ($\omega_n=2\pi c/\lambda_n$) and the width (full width at half maximum) of the resonance ($\Delta\omega$). If the frequency of the light is detuned by an amount $\delta\omega=\omega-\omega_n$, from the resonant frequency $\omega_n$ then the phase shift is $\theta-\theta_n=\delta\omega L/c$ where $L$ is the circumference of the ring. Then, since the shape of $P_{out}/P_{in}$ is inverse Lorentzian, we can express the same phase shift from resonance ($\theta-\theta_n$) in terms of $\Delta\omega$, where $\theta-\theta_n=\Delta\omega L/c$. Eq.~(\ref{eq:pout_original}) can be rewritten in terms of $\Delta\omega$ as
\begin{equation}
    P_{out}=\frac{(\frac{\alpha-t}{\sqrt{\alpha t}}\frac{c}{L})^2+(\Delta\omega)^2}{(\frac{1-\alpha t}{\sqrt{\alpha t}}\frac{c}{L})^2+(\Delta\omega)^2}P_{in}.
    \label{eq:pring_lor}
\end{equation}
Using the inverse Lorentzian shape of $P_{out}/P_{in}$, we obtain its spectral width when the two terms in the denominator of Eq.~(\ref{eq:pring_lor}) are equal, giving
\begin{equation}
    \Delta\omega=\frac{1-\alpha t}{\sqrt{\alpha t}}\frac{c}{L}.
\end{equation}
Since the resonance frequency is $\omega_n=2\pi c/\lambda_n$ and $L=2\pi R$, we express the loaded quality factor as
\begin{equation}
    Q=\frac{(2\pi)^2 R\sqrt{\alpha t}}{\lambda_n (1-\alpha t)}.
    \label{eq:Q}
\end{equation}
It is important to mention here that $Q$ diverges when $\alpha t\rightarrow1$, and deviations from this limit will give a finite $Q$ value, as shown in Fig.~\ref{fig:Qcontour}. We can see that for a ring radius $R=10$mm and at $\lambda_n=1550$nm, we can have a ring resonator with a loaded quality factor up to $Q\sim 2.5\times 10^7$ ($\alpha=1$, $t=0.99$).

\subsubsection{Phase Sensitivity}
We would like to explore the phase sensitivity of the field for the values of $t$ and $\alpha$ near one. We assume that the acquired phase shift $\phi$ is the smallest parameter in the problem to be sensitive to small rotations. We define the deviation parameters $\delta t=1-t$, $\delta\alpha=1-\alpha$, and $\delta\alpha,\delta t\ll1$ such that we will assume them to be small parameters in an expansion. Substituting these parameters, and $|(\theta\pm\phi)|\ll\delta\alpha,\delta t$ into Eq.~(\ref{eq:electricfield}), and expanding the equation, we find the resulting electric fields to be
\begin{multline}
    E_{t1}^{L,R}=\frac{1}{\delta\alpha+\delta t}\Bigg\{\delta\alpha-\delta t\\+\frac{2\delta t}{\delta\alpha+\delta t}[-i(\theta\pm\phi)-\frac{1}{2}(\theta\pm\phi)^2+...]\Bigg\} E_{i1}^{L,R}.
    \label{eq:Elong}
\end{multline}
In the limit where $|(\theta\pm\phi)|\ll\delta\alpha,\delta t\ll 1$ we can express this as
\begin{equation}
    E_{t1}^{L,R}\approx\bigg(\frac{\delta\alpha-\delta t}{\delta\alpha+\delta t}\bigg)e^{-i\Theta_{L,R}}E_{i1}^{L,R},
    \label{eq:phaseshift}
\end{equation}
where
\begin{equation}
    \Theta_{L,R}=-(\theta\mp\phi)\frac{2\delta t}{(\delta\alpha)^2-(\delta t)^2}.
    \label{eq:amp phase}
\end{equation}
In the highly overcoupling limit ($\delta\alpha\ll\delta t\ll1$), we interpret this expression as a phase amplification by a factor of $2/\delta t$ because of the cavity resonance effect.  The factor of 2 takes into account the phase shift difference between two opposite directions. The number of roundtrips the light takes around the integrated ring resonator gyroscope is given by $(cQ/\omega_n)/(2\pi R)$\cite{meitingthesis}. In highly overcoupled limit, this corresponds to $1/\delta t$, which is taken into account by Eq.~(\ref{eq:amp phase}).

When the system is on-resonance ($\theta=0$) and there is no rotation ($\phi=0$), following Eq.~(\ref{eq:Pout}) and (\ref{eq:Elong}), the power coming out of the ring in highly overcoupled limit is
\begin{equation}
    P_{out}=\bigg(\frac{\delta\alpha-\delta t}{\delta\alpha+\delta t}\bigg)^2P_{in}.
    \label{eq:prelation}
\end{equation}
Here we can see that the input power amplitude initially injected into the IWVA Sagnac interferometer is lost by a factor of $(\delta\alpha-\delta t)^2/(\delta\alpha+\delta t)^2$ after coming out of the ring resonator. 

The choice of the overcoupling limit as the optimal limit for our system can be justified based on the phase amplification given in Eq.~(\ref{eq:amp phase}). In critical limit ($\alpha=t$), there is no power coming out of the ring since $P_{out}=0$, and we cannot detect any power, as shown in Eq.~(\ref{eq:Pout}). In the undercoupling limit ($\delta t\ll\delta\alpha\ll1$), the phase is amplified by a finite factor $2\delta t/(\delta\alpha)^2$. In the overcoupling limit ($\delta\alpha\ll\delta t\ll 1$), the phase amplification is by a factor of $2/\delta t$, which is finite and much higher than the undercoupling limit. Moreover, all the power is emitted from the optical ring since $P_{out}\approx P_{in}$ for a highly overcoupled ring. For these reasons, we operate in the overcoupling limit.

\section{Results}\label{results}
In order to evaluate the performance of our system (IWVA Sagnac interferometer), we can compare it with the performance of a common Sagnac interferometer coupled to the ring resonator. For this comparison, we will look into the signal-to-noise ratios and phase resolutions of both systems.
\subsection{Phase Resolution - Common Sagnac Interferometer}
\begin{figure}[!htb]
\centering
 \includegraphics[scale=0.43]{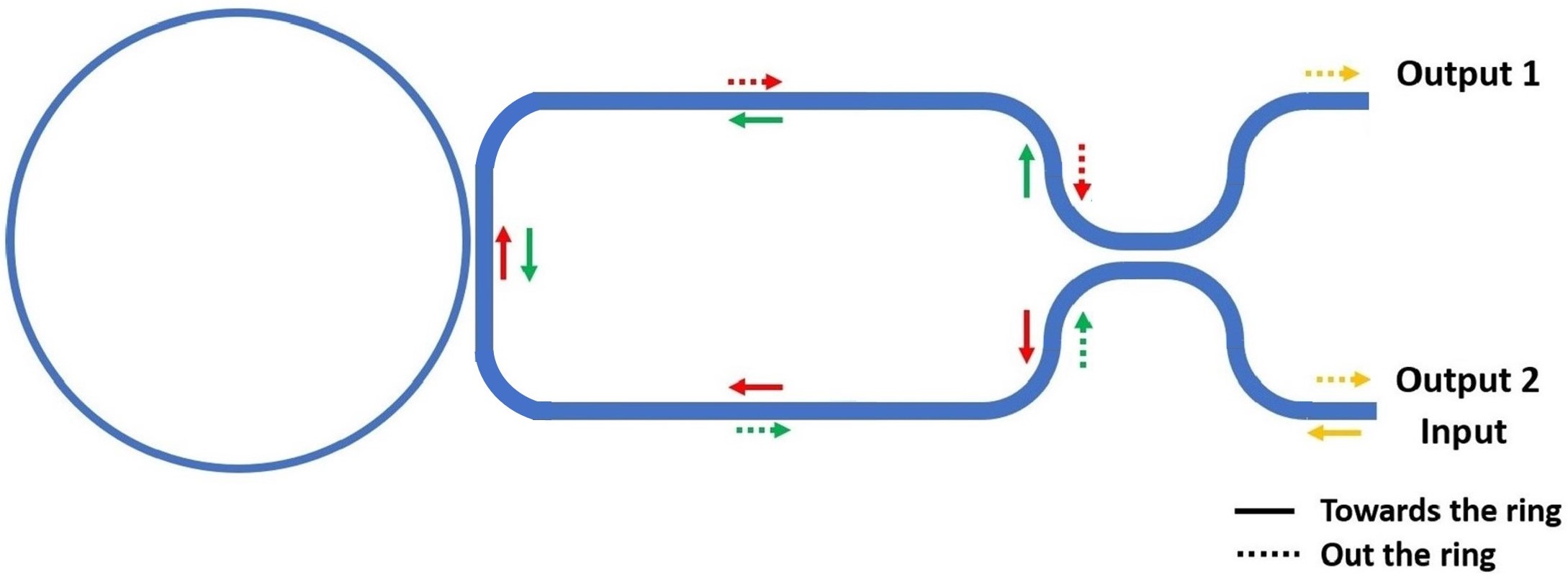}
 \caption{The considered device geometry (not to scale) of the common Sagnac interferometer coupled to the rotation sensing ring (left). Solid (dashed) arrows represent the input (output) field.}
 \label{fig:MZI}
\end{figure}

We start our analysis with the common Sagnac interferometer coupled to the ring resonator, as shown in Fig.~\ref{fig:MZI}. After the electric field loss and phase shift due to coupling with the ring, we can approximate the electric fields returning to the Sagnac interferometer from the ring resonator from both directions as
\begin{equation}
    E^{L}=\frac{1}{\sqrt{2}}\bigg(\frac{\delta\alpha-\delta t}{\delta\alpha+\delta t}\bigg)E_{in}(x,z)e^{-i\Theta_{L}}
    ,
    \label{eq:EtotMZI1}
\end{equation}
\begin{equation}
    E^{R}=\frac{i}{\sqrt{2}}\bigg(\frac{\delta\alpha-\delta t}{\delta\alpha+\delta t}\bigg)E_{in}(x,z)e^{-i\Theta_{R}}
    ,
    \label{eq:EtotMZI2}
\end{equation}
based on the relations we have provided in Eq.~(\ref{eq:prelation}) and (\ref{eq:amp phase}). Both Sagnac interferometer and IWVA Sagnac interferometer are assumed to be rectangular waveguides described in Section \ref{sub:input}. Therefore, the input electric field is the $\rm TE_0$ mode. Thus, $E_{in}(x,z)={\rm TE_0}(x,z)$. For a Sagnac interferometer with 50/50 beam splitters, the intensity output in both arms are given by $I_{1,2}=|(iE^L\mp E^R)/\sqrt{2}|^2$. Under the assumption of resonant coupling ($\theta=0$) and an intentional phase bias of $\pi/2$\cite{meitingthesis}, we can express this as
\begin{equation}
    I_{1,2}=\frac{I_0}{2}(1\pm {\rm sin}(2C\phi)),
\end{equation}
where $I_0=|E_0|^2$, $E_0=E_{in}(\delta\alpha-\delta t)/(\delta\alpha+\delta t)$, and $C=2\delta t/((\delta\alpha)^2-(\delta t)^2)$ is the phase amplification. Since we assume that the phase shift is small, we can use small angle approximation ${\rm sin}(2C\phi)\approx 2C\phi$ and take the difference of the two outputs as our measured signal, the signal per photon detected is expressed as 
\begin{equation}
    S=\frac{I_2-I_1}{I_1+I_2}=-2C\phi,
    \label{eq:signal standard}
\end{equation}
which yields a linear response in the phase $\phi$. Consequently, the signal-to-noise ratio (SNR) $\mathcal{R}$ is given by
\begin{equation}
    \mathcal{R}=\frac{N S}{\sigma}
\end{equation}
where $N$ is the number of photons injected into the system and detected, and $\sigma$ is the standard deviation (noise) of the two photon detectors measuring $I_1$ and $I_2$. Since the source of the noise for coherent laser light is the shot-noise, and the two detectors are uncorrelated, the variance is expressed as $\sigma^2[I_1-I_2]=\sigma^2[I_1]+\sigma^2[I_2]=N_1+N_2$, where $N_1$ and $N_2$ are the number of photons detected at the two outputs, and $N_1+N_2=N$. Hence, the standard deviation is $\sigma=\sqrt{N}$. In Sec.~\ref{subsec allan}, we do an analysis beyond the shot noise limit and discuss other potential sources of error. The use of quantum light, such as squeezed light can further reduce these fluctuations. Then, $\mathcal{R}$ can be expressed as 
\begin{equation}
    \mathcal{R}=\sqrt{N}\frac{(-4)\delta t}{(\delta\alpha)^2-(\delta t)^2}\phi.
    \label{eq:R}
\end{equation}
 At the overcoupling limit where $\delta\alpha\ll\delta t\ll 1$, we can simply assume that $\delta t$ is small and finite and $\delta\alpha\rightarrow0$. Then we are left with $\mathcal{R}=4\sqrt{N}\phi/\delta t$, which is the largest SNR we can get. In overcoupling limit, the loaded quality factor provided in Eq.~(\ref{eq:Q}) will be $Q=(2\pi)^2R/(\delta t\lambda_n)$. Therefore, assuming that the ring is lossless, we can express this ideal SNR in overcoupling limit in terms of $Q$ as
\begin{equation}
    \mathcal{R}=\frac{\sqrt{N}Q\lambda_n}{\pi^2R}\phi.
    \label{eq:R_final}
\end{equation}
Since a signal producing an SNR of unity indicates the smallest practically resolvable signal\cite{barnett2003ultimate,dressel2013strengthening}, we can find the smallest detectable phase by setting $\mathcal{R}=1$, to find
\begin{equation}
    \phi_{min}=\frac{\pi^2R}{\sqrt{N}Q\lambda_n}.
\end{equation}
The minimal angular rotation rate that can be detected can be found from this phase. Via the Sagnac relation\cite{passiveringres}, we know that the angular rotation rate is $\Omega=\ell c/(2\pi R^2)$, where $c$ is the speed of light, and via the geometric phase, we know that $\ell=\phi\lambda_n/(2\pi)$. The variables in these relations are defined in Sec.~\ref{subsec:C}. Hence, the minimal detectable angular rotation rate is given by
\begin{equation}
    \Omega_{min}=\frac{c}{4RQ\sqrt{N}}.
    \label{eq:MZI_omg_min}
\end{equation}
From this equation, we see that we can minimize the angular rotation rate by maximizing the product of the radius of the ring ($R$) and the loaded quality factor ($Q$), for a fixed photon number.

\subsection{Phase Resolution - IWVA Sagnac Interferometer}
We can now analyze the minimal angular rotation rate for the IWVA Sagnac interferometer discussed in this paper and compare it with the Sagnac interferometer discussed in previous section. The electric fields returning to the IWVA Sagnac interferometer from the ring resonator from both directions are expressed the same as in Eq.~(\ref{eq:EtotMZI1})  and (\ref{eq:EtotMZI2}), and with the same definition for $E_{in}(x,z)$. As the electric field in each waveguide propagates through the interferometer, it goes through a spatial phase front tilter where a portion of the light in $\rm TE_0$ mode is tapped out and then reinjected to the broadened waveguide as $\rm TE_1$ mode\cite{song2021enhanced,steinmetz2022enhanced,steinmetz2019precision}, as shown in Fig.~\ref{fig:setup}{\color{red}b} and discussed in Sec.~\ref{sec2}. Hence, the initial electric field mode transforms into 
\begin{equation}
    {\rm TE_0}(x,z)\rightarrow\sqrt{1-a^2}{\rm TE_0}(x,z)\pm ia{\rm TE_1}(x,z)
\end{equation}
for the left and right waveguide arms. Here, $a$ is the amplitude of the $\rm TE_0$ mode that was tapped out and coupled to the $\rm TE_1$ mode\cite{song2021enhanced,steinmetz2022enhanced,steinmetz2019precision}. Then, modifying the input electric field for Sagnac interferometer, given in Eq.~(\ref{eq:EtotMZI1}) and (\ref{eq:EtotMZI2}), we can express the electric field in each waveguide of the IWVA Sagnac interferometer as
\begin{equation}
  E^{L}=\frac{e^{-i\Theta_{L}}}{\sqrt{2}}\bigg(\frac{\delta\alpha-\delta t}{\delta\alpha+\delta t}\bigg)[\sqrt{1-a^2}{\rm TE_0}(x,z)+ ia{\rm TE_1}(x,z)],
    \label{eq:ELtaper} 
\end{equation}
and
\begin{equation}
  E^{R}=\frac{e^{-i\Theta_{R}}}{\sqrt{2}}\bigg(\frac{\delta\alpha-\delta t}{\delta\alpha+\delta t}\bigg)[i\sqrt{1-a^2}{\rm TE_0}(x,z)+a{\rm TE_1}(x,z)].
    \label{eq:ERtaper} 
\end{equation}
The electric fields in both waveguides will then propagate through the directional coupler we initially used for the input field. We can treat the ${\rm TE_0}$ mode and the ${\rm TE_1}$ mode of the total electric field going into the coupler separately such that the $\rm TE_0$ modes of both waveguides will have the coupling constant $\kappa^{\rm DC}_0$ defined in Sec.~\ref{sec2} and the $\rm TE_1$ modes of both waveguides will have a different coupling constant $\kappa^{\rm DC}_1=\kappa^{\rm DC}_{LR}=\kappa^{\rm DC}_{RL}$ based on Eq.~(\ref{eq:couplingconstant}) where $E^L$ and $E^R$ are in ${\rm TE_1}$ mode only. Since $\kappa^{\rm DC}_1$ is not necessarily the same as $\kappa^{\rm DC}_0$, the $\rm TE_0$ and $\rm TE_1$ modes in both arms will have different coupling constants. Thus, to achieve a 50/50 beam splitter, the coupling length $S$ needed for the two modes tends to be different. However, we can benefit from the periodicity of the coupling process and design the length of the directional coupler such that the $\rm TE_0$ mode goes through 1/8 (1/4) of a coupling cycle of the electric field (intensity), while the $\rm TE_1$ mode goes through 9/8 (5/4) of a coupling cycle of the electric field (intensity)\cite{song2021enhanced}. Hence, we obtain a 50/50 beam splitter for both modes simultaneously. 

As the beams propagate through the multimode 50/50 beam splitter and interfere before being detected, we obtain two output fields which we call as the ``bright'' and ``dark" modes where 
\begin{align}
    E_b=\frac{iE^L+E^R}{\sqrt{2}},\\
    E_d=\frac{iE^L-E^R}{\sqrt{2}}.
\end{align}
Under the assumptions $|C\phi|\ll1$ and $a\ll1$, we can express the bright mode as
\begin{equation}
    E_b(x)=\bigg(\frac{\delta\alpha-\delta t}{\delta\alpha+\delta t}\bigg)i\bigg[{\rm TE_0}(x)-aC\phi{\rm TE_1}(x)\bigg],
\end{equation} and the dark mode as
\begin{equation}
    E_d(x)=\bigg(\frac{\delta\alpha-\delta t}{\delta\alpha+\delta t}\bigg)a\bigg[\frac{C\phi}{a}{\rm TE_0}(x)-{\rm TE_1}(x)\bigg]. 
    \label{eq:darkport}
\end{equation}
It is important to note that in the bright mode, the phase information is carried by the $\rm TE_1$ mode, and is suppressed by a factor of $a\ll 1$. In the dark mode, after renormalization, we have mainly a $\rm TE_1$ mode with a small amount of $\rm TE_0$ mode added in\cite{steinmetz2022enhanced}. Unlike the bright mode, the phase information is carried by the $\rm TE_0$ mode in the dark mode, and yields an inverse weak value amplification by a factor of $1/a\gg 1$. However, in the high overcoupling limit, the overall input field is reduced by a factor of $a$. Since the phase information is suppressed in bright mode, and amplified in dark mode, we post-select the dark mode for phase measurements. We can then measure the ratio between the modes $\rm TE_0$ and $\rm TE_1$ by using a multimode interferometer (MMI) at the dark port which has an output power that depends on the mode ratio as discussed in Refs.\cite{steinmetz2022enhanced,steinmetz2019precision}, to obtain the measured signal
\begin{equation}
    S_{IWVA}=-\frac{C}{a}\phi,
    \label{eq:signal WVA}
\end{equation}
where $C=2\delta t/((\delta\alpha)^2-(\delta t)^2)$ is the phase amplification discussed in previous section. While other methods have been discussed in Refs.\cite{steinmetz2022enhanced,steinmetz2019precision}, it has been concluded that the mode ratio method is the optimal one and hence, we will use this method in this paper. Similar to the Sagnac interferometer case (Eq.~(\ref{eq:R})), under the assumption $\delta\alpha\ll\delta t$, the SNR of the IWVA Sagnac interferometer can be expressed as
\begin{equation}
    \mathcal{R}_{IWVA}=\frac{4\sqrt{N_{PS}}}{\delta t}\frac{1}{2a}\phi,
\end{equation}
where $N_{PS}$ is the number of photons detected after post-selection, which is less than the number of injected photons. $\mathcal{R}_{IWVA}$ can be expressed in terms of $R$ and $Q$ just like in the Sagnac interferometer case (Eq.~(\ref{eq:R_final}))
\begin{equation}
    \mathcal{R}_{IWVA}=\frac{\sqrt{N_{PS}}Q\lambda_n}{\pi^2R}\frac{\phi}{2a}.
    \label{RMMI}
\end{equation}
Finally, the minimal angular rotation rate can be derived the same way as in the Sagnac case in Eq.~(\ref{eq:MZI_omg_min})
\begin{equation}
    \Omega_{min}^{'}=2a\frac{c}{4RQ\sqrt{N_{PS}}},
    \label{eq:MMIprecision}
\end{equation}which can be expressed in terms of the injected photon number $N_{in}^{IWVA}$
\begin{equation}
    \Omega_{min}^{'}=2\frac{c}{4RQ\sqrt{N_{in}^{IWVA}}},
    \label{eq:MMIprecision2}
\end{equation}
where $N_{PS}=|a|^2 N_{in}^{IWVA}$. We see that when the detected power is the same for the Sagnac interferometer and the IWVA Sagnac interferometer ($N=N_{PS}$), the minimal angular rotation rate using the IWVA Sagnac interferometer is more precise by a factor of $2a$ than using the Sagnac interferometer.

\subsection{Comparison - IWVA Sagnac Interferometer vs. Sagnac Interferometer}

\begin{figure}[!htb]
\centering
 \includegraphics[scale=0.35]{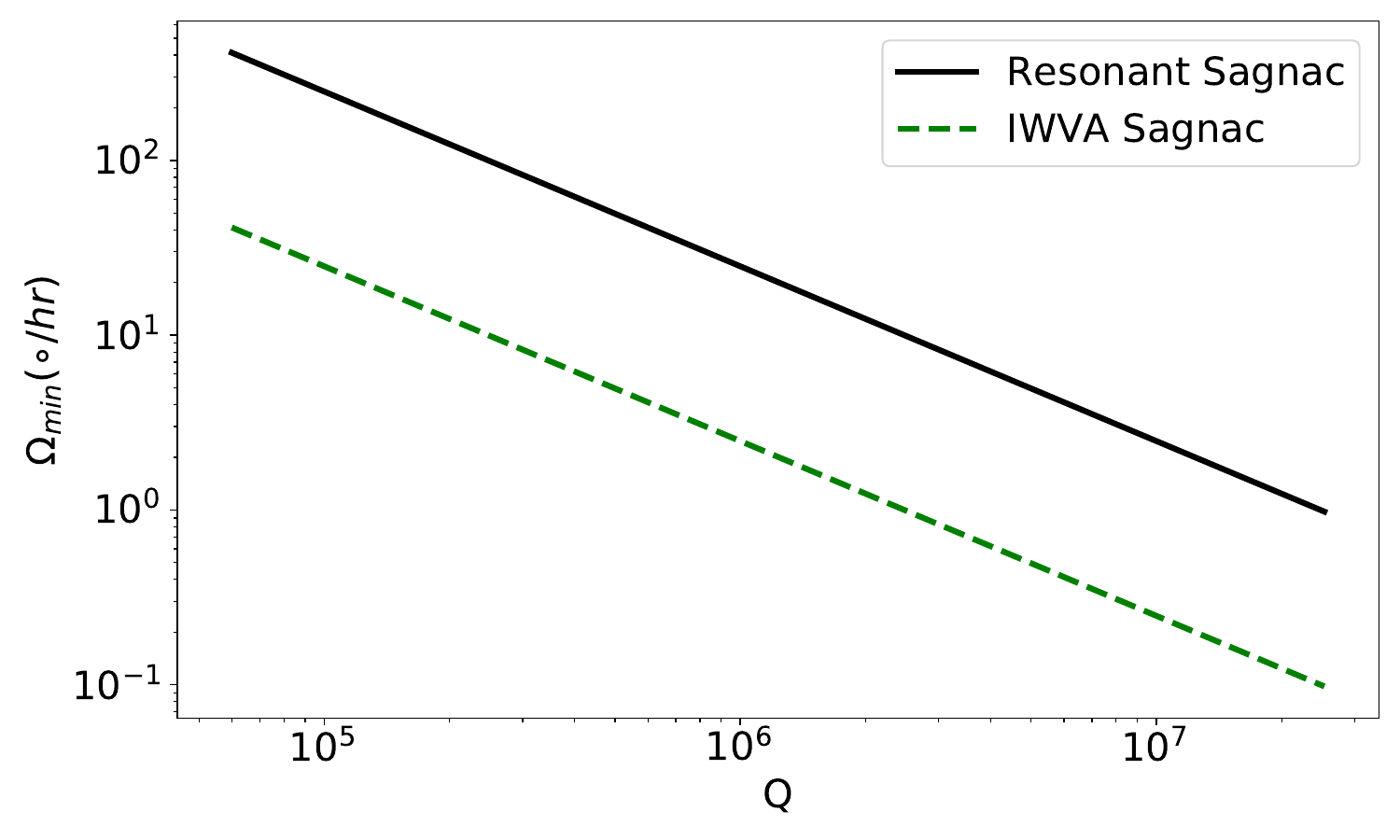}
 \caption{The angular rotation rate resolution for Sagnac interferometer (black solid line) and for IWVA Sagnac interferometer (green dashed line), in units of $^{\circ}/$hr are plotted against the loaded quality factor $Q$. Here we obtain the $Q$ values by taking $\alpha=1$ and t ranging from 0.05 to 0.99, satisfying the overcoupling limit. We assume an integration time of 1s, $R=10$mm, $a=0.05$, $\lambda_n=1550$nm, and optical power at the detector to be $P_{det}=0.5$mW for both systems ($N_{PS}=N$). For the Sagnac interferometer results, we also have a $0.5$mW input, but the inverse weak-value design has a $200$mW input, maintaining a $0.5$mW detected power which is the saturation power of the detectors in typical detectors.}
 \label{fig:resolutionplots}
\end{figure}

\begin{figure}[!htb]
\centering
 \includegraphics[scale=0.35]{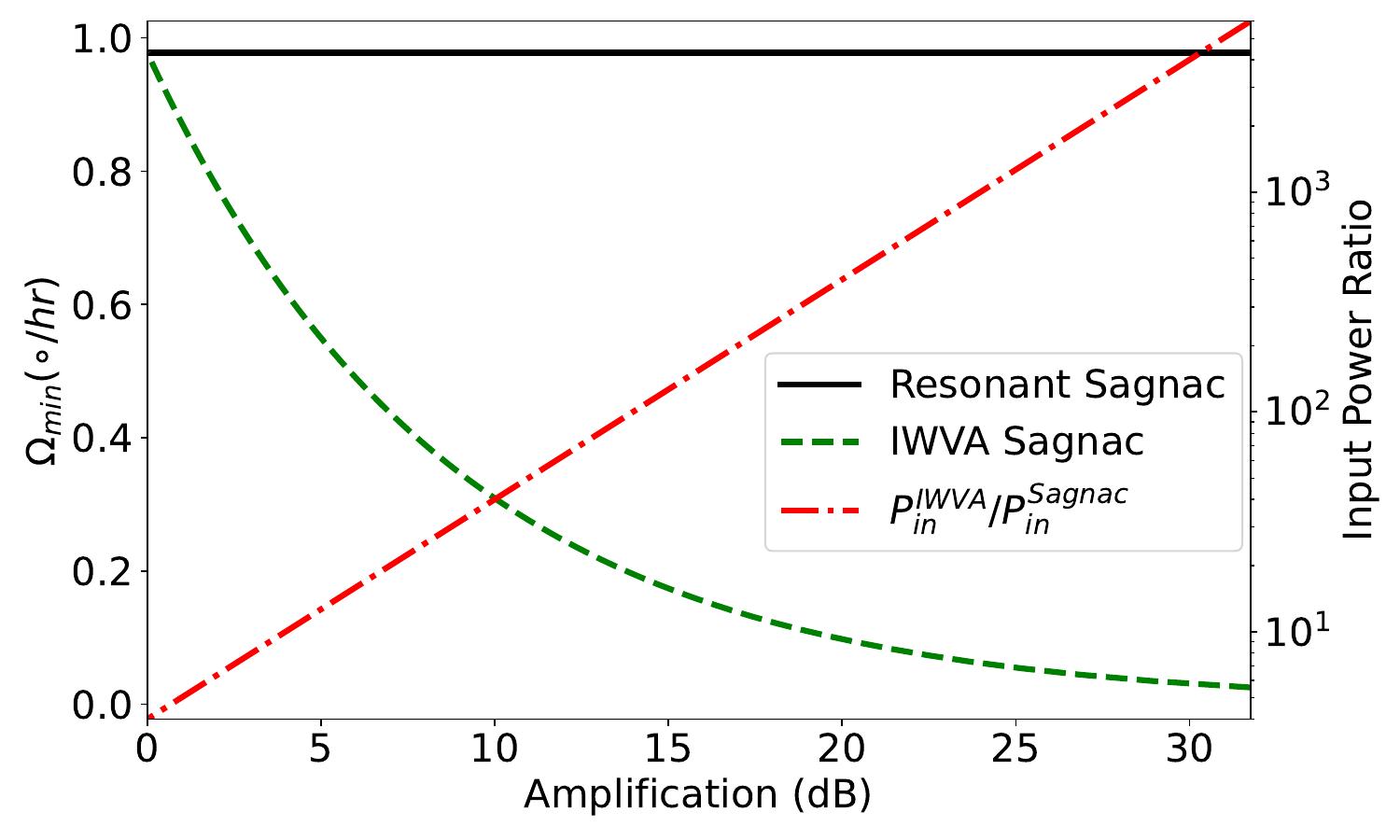}
 \caption{The angular rotation rate resolution (left y-axis) for Sagnac interferometer (black solid line) and for IWVA Sagnac interferometer (green dashed line), in units of $^{\circ}/$hr, and the input power ratio (right y-axis) between the IWVA Sagnac interferometer and Sagnac interferometer (red dash-dot line) are plotted against the phase amplification of the IWVA Sagnac interferometer relative to the Sagnac interferometer. We assume an integration time of 1s, $Q\sim 2.5\times10^7$ ($\alpha=1,$ $t=0.99$), $R=10$mm, $\lambda_n=1550$nm, and optical power at the detector to be $P_{det}=0.5$mW for both systems ($N_{PS}=N$).}
 \label{fig:sensvsamp}
\end{figure}
To see the impact of the inverse weak value amplification achieved with our system, it is important to note that the relation between $\mathcal{R}_{IWVA}$ and $\mathcal{R}$ is given by
\begin{equation}
    \frac{\mathcal{R}_{IWVA}}{\mathcal{R}}=\sqrt{\frac{N_{PS}}{N}}\frac{1}{2a}.
    \label{Rcomp}
\end{equation} Based on Eq.~(\ref{eq:darkport}), the power detected for the IWVA Sagnac interferometer is expressed as $P_{det}^{IWVA}=|a|^2P_{in}^{IWVA}$ where $P_{in}^{IWVA}$ is the input power of the IWVA Sagnac interferometer. Since the power and number of photons are directly proportional, the number of detected photons can be expressed as $N_{PS}=|a|^2 N_{in}^{IWVA}$ where $N_{in}^{IWVA}$ is the number of photons injected into IWVA Sagnac interferometer. On the other hand, the power detected for the Sagnac interferometer is the same as the power injected into it, $P_{det}^{Sagnac}=P_{in}^{Sagnac}$. Given that $a$ is real, the relation above can be expressed as
\begin{equation}
    \frac{\mathcal{R}_{IWVA}}{\mathcal{R}}=\frac{1}{2}\sqrt{\frac{N_{in}^{IWVA}}{N}}.
    \label{Rcomp2}
\end{equation}
Eq.~(\ref{eq:MMIprecision}), (\ref{Rcomp2}), and the direct proportionality between the power and the number of photons show that IWVA Sagnac interferometer has a better SNR and more precise angular rotation rate measurement than a Sagnac interferometer, given that $P_{in}^{IWVA}>4P_{in}^{Sagnac}$. Then the SNR and the minimum detectable angular rotation rate can both be improved 10 times for $P_{in}^{IWVA}=400P_{in}^{Sagnac}$. Therefore, we can boost the input power of IWVA Sagnac interferometer to make it perform better than a Sagnac interferometer, at the same detected power value.

We then compare the simulated angular rotation rate resolutions ($\Omega_{min}$) as a function of the loaded quality factor $Q$ in the overcoupling limit where $\alpha=1$ and $t$ ranges from 0.05 to 0.99 for IWVA Sagnac interferometer (green dashed line) and Sagnac interferometer (black solid line) in Fig.~\ref{fig:resolutionplots}. We take $R=10$mm, $\lambda_n=1550$nm, $a=0.05$, and the input power to be $P_{in}=0.5$mW for the Sagnac interferometer, and $P_{in}=200$mW for the IWVA Sagnac interferometer in overcoupling limit $\delta\alpha\ll\delta t$. The integration time is taken to be 1s. In both cases we have the same detected power of $P_{det}=0.5$mW which corresponds to a postselection probability of $|a|^2=1/400$ for the IWVA Sagnac interferometer. The postselection probability can be further reduced in fabrication. This detected power also corresponds to the saturation power of typical detectors. We see that we can gain a factor of 10 or more in precision just from the ability to use a higher power, but keeping the detected power at the saturation point. 

In Fig.~\ref{fig:sensvsamp}, we show the angular rotation rate resolution (on the left y-axis) for the Sagnac (solid black line) and the IWVA Sagnac (green dashed line), against the amplification (in dB) of the IWVA Sagnac phase relative to the Sagnac phase (20 $\rm log_{10}(1/2a)$). Compared to Fig.~\ref{fig:resolutionplots}, we vary the amplification while keeping the loaded quality factor fixed at $Q\sim 2.5\times 10^7$ ($\alpha=1,$ $t=0.99$). The Sagnac phase is not amplified. Therefore, it is constant, whereas the IWVA Sagnac phase resolution decays below $0.1^\circ/\rm hr$ for amplification beyond 20 dB. On the right y-axis, we evaluate the ratio between the input power of the IWVA Sagnac relative to the Sagnac, which is fixed at $0.5$ mW. The input power required for the IWVA Sagnac increases linearly with amplification in log-log scale. This reiterates our previous point: improving the precision of the angular rotation rate comes at the cost of increased power input that has to be injected into the system.

For both Figs.~\ref{fig:resolutionplots} and \ref{fig:sensvsamp}, under expected ideal conditions, for a loaded quality factor $Q\sim2.5\times10^7$ ($\alpha=1,$ $t=0.99$) and radius $R=10$mm, we have a precision of less than $0.1^{\circ}/$hr for IWVA Sagnac interferometer, which exceeds the current state of the art\cite{GUO2012302,dell2014recent,ciminelli2010photonic,ciminelli2012high}. Even better precision can be obtained by fabricating a ring resonator with larger radius as we suggest in Eq.~(\ref{eq:MMIprecision}). Given that an intrinsic $Q$ of 720 million\cite{liu2022ultralow} and 422 million\cite{puckett2021422} has been achieved for a $\rm Si_3N_4$ bus-coupled ring resonator in the past, we can obtain even better rotation sensitivity.

\subsection{Allan Variance}
\label{subsec allan}
\begin{figure}[!htb]
\centering
 \includegraphics[scale=0.35]{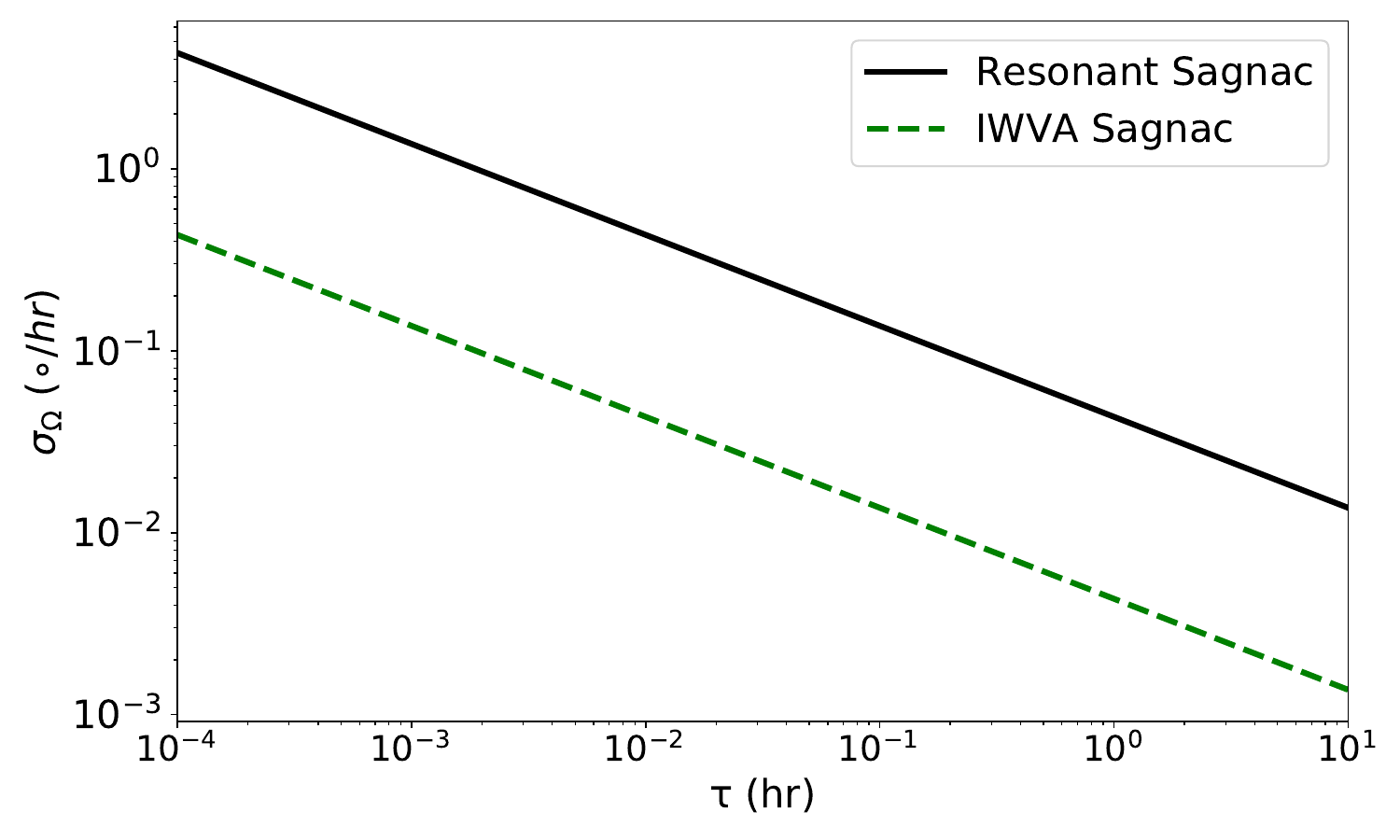}
 \caption{Allan deviation for the Sagnac interferometer (solid black line) and IWVA Sagnac interferometer (dashed green line) as a function of the averaging time $\tau$ in hours. We take $Q\sim2.5\times10^7$ ($\alpha=1,$ $t=0.99$), $\eta=0.8{\rm A/W}$, $T=293{\rm K}$, $R_f=100 {\rm ohms}$, $i_d=1{\rm nA}$, $R=10$mm, $a=0.05$, $\lambda_n=1550$nm, and optical power at the detector to be $P_{det}=0.5$mW for both systems.}
 \label{fig:ADEV}
\end{figure}
We finally want to model the uncertainty and error mechanics of the IWVA Sagnac interferometer and compare it to the Sagnac interferometer to see the improvement in error due to IWVA. The errors of these interferometers primarily stem from the noise generated by the measurement instruments, readout detectors. While there are numerous variance techniques available to measure the stochastic errors of inertial sensors, we will be using Allan variance which is one of the simplest and prevalent methods in the field\cite{el2007analysis,lefevre2022fiber}. For Allan variance calculation, we can assume $M$ consecutive data points, each with a time interval $t_0$. We can divide these samples into clusters, each with $n$ consecutive data points ($n<M/2$). Each cluster will have a total time interval of $\tau=nt_0$. Then, Allan variance ($\sigma_{\Omega}^2(\tau)$) is defined as

\begin{equation}
    \sigma_{\Omega}^2(\tau)=\frac{1}{2(M-2n)}\sum_{k=1}^{M-2n}[\overline{\Omega}_{k+1}(\tau)-\overline{\Omega}_{k}(\tau)]^2,
\end{equation}
where $k$ is the index for the cluster number measured in the time interval between $t_k$ and $t_k+\tau$, and $\overline{\Omega}_{k}(\tau)=\frac{1}{\tau}\int_{t_k}^{t_k+\tau} \Omega(t) \,dt $ is the time averaged angular rotation rate for a given cluster $k$.
Based on signal processing theory, Allan variance can also be expressed as the integral of the product of the power spectral density ($PSD_\Omega$) of the noise in the detector by the square of the Fourier transform of the
time gating function over a duration $\tau$\cite{el2007analysis}

\begin{equation}
    \sigma_{\Omega}^2(\tau)=4\int_{0}^{\infty} PSD_\Omega(f)\frac{{\rm sin}^4(\pi f\tau)}{(\pi f\tau)^2} \,df,
\end{equation}
where $f=\omega/2\pi$ is the frequency and $PSD_\Omega$ has the units $(^{\circ}/\rm hr)^2/\rm Hz$. In this work we assume that both Sagnac and IWVA Sagnac are white noise limited. This is a reasonable assumption for short measurement time $\tau$ where white noise is the dominant noise source. Hence, the power spectral density will be independent of the frequency $f$\cite{660628}, and we can express {the Allan deviation ($\sigma_{\Omega}(\tau)$)} in terms of the angle random walk coefficient ($\sigma_{\theta}$)\cite{gu2013random}
\begin{equation}
    \sigma_{\Omega}(\tau)=\frac{\sigma_{\theta}}{\sqrt{\tau}},
\end{equation}
where $\sigma_{\theta}=\sqrt{PSD_\Omega}$, but with units $^{\circ}/\sqrt{\rm hr}$. Despite both $\sigma_{\theta}$ and $\sqrt{PSD_\Omega}$ have the same dimensions, they have different SI units. Hence, the unit conversion from $\sqrt{PSD_\Omega}$ to $\sigma_{\theta}$ is $1^{\circ}/\sqrt{\rm hr}=60(^{\circ}/\rm hr)/\sqrt{\rm Hz}$\cite{gu2013random}.

As discussed in Refs.\cite{li2014multi,blake1997random}, the noise generated by the optical detectors directly affect the random walk coefficient ($\sigma_{\theta}$) of the gryroscope and the noise limit is the detection noise\cite{lefevre2022fiber}. Therefore, $\sigma_{\theta}$ stems from the noise of the detectors used for the angular rotation readout. Then, the power spectral density can be expressed as

\begin{equation}
    PSD_\Omega(\omega)=\int_{-\infty}^{\infty} \langle\Omega(t)\Omega(t+t')\rangle e^{-i\omega t} \,dt'.
\end{equation}
We know from Sagnac relation, discussed in Sec.~\ref{subsec:C}, that $\phi$ is the only time dependent variable in detection. Then,
\begin{equation}
    \Omega(t)=\frac{\lambda_n c}{4\pi A}\phi(t),
\end{equation}
where $A=\pi R^2$ is the area of the ring resonator. Hence, we can rewrite the power spectral density as
\begin{equation}
    PSD_\Omega(\omega)=\bigg(\frac{\lambda_n c}{4\pi A}\bigg)^2\int_{-\infty}^{\infty} \langle\phi(t)\phi(t+t')\rangle e^{-i\omega t} \,dt'.
\end{equation}
For $t''=t+t'$, and assuming in short time scale for $\tau$ the phase $\phi$ is a delta-correlated random variable, we can express the autocorrelation function as
\begin{equation}
    \langle\phi(t)\phi(t'')\rangle=\delta(t-t'')\langle\phi^2\rangle.
\end{equation}
Since the mean is zero for white noise, $\langle\phi\rangle=0$. Given that the deviation from average is defined as $\langle\delta\phi\rangle=\phi-\langle\phi\rangle$, we can write the standard deviation as $\langle\delta\phi^2\rangle=\langle\phi^2\rangle-\langle\phi\rangle^2=\langle\phi^2\rangle$. Then the autocorrelation function can be written as
\begin{equation}
    \langle\phi(t)\phi(t'')\rangle=\delta(t-t'')\langle\delta\phi^2\rangle,
\end{equation}
and the power spectral density becomes
\begin{equation}
    PSD_\Omega(\omega)=\bigg(\frac{\lambda_n c}{4\pi A}\bigg)^2\langle\delta\phi^2\rangle.
\end{equation}
This also shows that Allan deviation is equal to standard deviation when it is white noise limited\cite{lefevre2022fiber}. Finally, $\sigma_{\theta}$ is expressed as
\begin{equation}
    \sigma_{\theta}=\frac{\lambda_n c}{4\pi A}\sqrt{\langle\delta\phi^2\rangle}.
\end{equation}
It is worth noting that in the overcoupling limit, the detected phase is amplified by $-2C$ and $-C/a$ for the Sagnac and IWVA Sagnac respectively. The standard deviation $\langle\delta\phi^2\rangle$ represents the total noise that stems from Johnson, shot and dark current noise\cite{blake1997random,li2014multi}. Taking these factors into account, and normalizing the Sagnac phase for amplification, in the overcoupling limit, $\sigma_{\theta}$ for the Sagnac is given by
\begin{equation}
    \sigma_{\theta}^{Sagnac}=\frac{\lambda_n c}{4\pi A\eta P_{det}}\bigg(\frac{\delta t}{4}\bigg)\sqrt{\frac{4k_B T}{R_f}+2e\eta P_{det}+2ei_d},
\end{equation}
and $\sigma_{\theta}$ for IWVA Sagnac is
\begin{equation}
    \sigma_{\theta}^{IWVA}=\frac{\lambda_n c}{4\pi A\eta P_{det}}\bigg(\frac{a\delta t}{2}\bigg)\sqrt{\frac{4k_B T}{R_f}+2e\eta P_{det}+2ei_d},
\end{equation}
where $P_{det}=P_{det}^{Sagnac}=P_{det}^{IWVA}$ is the power detected as defined before, $\eta$ is the detector efficiency in units of $\rm A/W$, $e$ is the electron charge, $k_B$ is the Boltzmann constant, $T$ is the temperature, $R_f$ is the detector trans-impedance, and $i_d$ is the dark current. Defining an effective area $A_{eff}=\lambda_n RQ/\pi$ for the ring resonator, where $Q=(2\pi)^2R/(\delta t\lambda_n)$, we can express the Allan deviation (also standard deviation in white noise limit) as
\begin{multline}
    \sigma_{\Omega}^{Sagnac}(\tau)=\frac{\lambda_n c}{4\pi A_{eff}\eta P_{det}
    \sqrt{\tau}} \\
    \times\sqrt{\frac{4k_B T}{R_f}+2e\eta P_{det}+2ei_d},
\end{multline}
\begin{multline}
    \sigma_{\Omega}^{IWVA}(\tau)=2a\frac{\lambda_n c}{4\pi A_{eff}\eta P_{det}\sqrt{\tau}}\\
    \times\sqrt{\frac{4k_B T}{R_f}+2e\eta P_{det}+2ei_d}.
\end{multline}

For $Q\sim2.5\times10^7$ ($\alpha=1,$ $t=0.99$), $\eta=0.8{\rm A/W}$, $T=293{\rm K}$, $R_f=100 {\rm ohms}$, $i_d=1{\rm nA}$, $\tau=10{\rm s}$, we find $\sigma_{\Omega}^{Sagnac}=0.82^{\circ}/\rm hr$ and $\sigma_{\Omega}^{IWVA}=0.08^{\circ}/\rm hr$. We also note that the thermal and shot noise are of the same order at $1.6\times10^{-22}$ and $1.3\times10^{-22} \rm A^2s$ and much more dominant than the dark current noise at $3.2\times10^{-28} \rm A^2s$. Other parameters have been specified in previous sections. Improved precision in angular rotation rate readouts as a result of the inverse weak value amplification provides an Allan deviation within an order of magnitude of $10^{-2}(^{\circ}/\rm hr)$, and improves the Allan deviation by more than ten times compared to Sagnac interferometer. For a wider range of averaging time $\tau$, we can see in Fig.~\ref{fig:ADEV} that the Allan deviations for both systems have the same negative slope decays in log-log scale\cite{lefevre2022fiber}, but the IWVA Sagnac provides over ten times improvement over the Sagnac at any time.

\section{Conclusions}\label{conc}
In this work, we have shown that our IWVA Sagnac interferometer allows us to amplify the Sagnac phase shift by a factor of $1/a$, using only a small fraction of the input power. Compared to a Sagnac interferometer, IWVA Sagnac interferometer has a potential to significantly increase the signal-to-noise ratio and the phase resolution, and reduce the Allan deviation, given that the power detected after post-selection is at the detector saturation point and comparable enough to the power detected at the Sagnac interferometer. This is done through injecting a much larger input power into the IWVA Sagnac interferometer compared to the Sagnac interferometer. We have shown in Fig.~\ref{fig:resolutionplots} that when we inject a larger input power to IWVA Sagnac interferometer and detect the same amount of power as Sagnac interferometer, we can improve the precision of the minimal angular rotation rate by more than a factor of 10. Using Eq.~(\ref{Rcomp}), we can also see that under these parameters, we can improve the SNR by more than a factor of 10. Similarly, errors in detection can be reduced by more than ten times. Moreover, for ideal parameters, we obtain a minimum detectable angular rotation rate $\Omega_{min}^{'}\sim0.1^{\circ}/$hr and Allan deviation $\sigma_{\Omega}^{IWVA}\sim0.08^{\circ}/$hr, which enables applications in aerospace and defense industry. Inverse weak value amplification with IWVA Sagnac interferometer developed in this paper can also have potential applications in laser frequency stabilization and other applications in metrology in the future.
\begin{acknowledgments}
This work was supported by NSF award ECCS: 2330328 and Leonardo DRS Technologies.
\end{acknowledgments}

\bibliography{references}

\end{document}